# Domain Wall Propagation and Pinning Induced by Current Pulses in Cylindrical Modulated Nanowires

C.Bran[1], J.A. Fernandez-Roldan[2], J. A. Moreno[3], A. Fraile Rodríguez[4,5], R. P. del Real[1], A. Asenjo[1], E. Saugar[1], J. Marqués-Marchán[1], H. Mohammed[3], M. Foerster[6], L. Aballe[6], J. Kosel[3,7], M. Vazquez[1], O. Chubykalo-Fesenko[1]

[1] Instituto de Ciencia de Materiales de Madrid, 28049 Madrid, Spain

[2] Helmholtz-Zentrum Dresden-Rossendorf e.V., Institute of Ion Beam Physics and Materials Research, Bautzner Landstrasse 400, 01328 Dresden, Germany

[3] King Abdullah University of Science and Technology, Computer Electrical and Mathematical Science and Engineering, Thuwal 23955-6900, Saudi Arabia.

[5] Departament de Física de la Matèria Condensada, Universitat de Barcelona, Barcelona, 08028, Spain

Institut de Nanociencia i Nanotecnologia (IN2UB), Universitat de Barcelona, Barcelona, 08028, Spain

[6] ALBA Synchrotron Light Facility, CELLS, Barcelona, 08290, Spain

[7] Sensor Systems Division, Silicon Austria Labs, Villach 9524, Austria

**Abstract**

The future developments of three-dimensional magnetic nanotechnology require the control of domain wall dynamics by means of current pulses. While this has been extensively studied in planar magnetic strips (planar nanowires), few reports exist in cylindrical geometry, where Bloch point domain walls are expected to have intriguing properties. Here we report this investigation in cylindrical magnetic Ni nanowires with geometrical notches. Experimental work based on synchrotron X-ray magnetic circular dichroism (XMCD) combined with photoemission electron microscopy (PEEM) indicates that large current densities induce domain wall nucleation while smaller currents move domain walls preferably against the current direction. In the region where no pinning centers are present we found domain wall velocity of about 1 km/s. The domain wall motion along current was also detected in the vicinity of the notch region. Pinning of domain walls has been observed not only at geometrical constrictions but also outside of them. Thermal modelling indicates that large current densities temporarily raise the temperature in the nanowire above the Curie temperature leading to nucleation of domain walls during the system cooling. Micromagnetic modelling with spin-torque effect shows that for intermediate current densities Bloch point domain walls with chirality parallel to the Oersted field propagate antiparallel to the current direction. In other cases, domain walls can be bounced from the notches and/or get pinned outside their positions. We thus find that current is not only responsible for the domain wall propagation but is also a source of pinning due to the Oersted field action.

**Introduction**

Cylindrical magnetic nanowires provide versatile functionalities for data and energy storage, sensing, magnetic nanocircuits or magneto-mechanical actuators.[1] They are promising candidates as building blocks for novel three-dimensional nanotechnology.[2] The future implementation of such technology requires manipulation of magnetism in cylindrical magnetic nanowires by means of low power consumption stimuli such as electric currents. Spintronics is widely recognized within the scientific/technological community as candidate for future energy-saving nano-applications.[3] Comparatively to the use of external fields, current-induced magnetization dynamics offer more energy efficiency. However, unlike planar nanowires, spintronic-based manipulation of magnetism in cylindrical magnetic nanowires has not yet been developed despite their high potential for high storage density and other novel multifunctionalities.

Magnetic DWs are expected to play a decisive role as information carriers in magnetic circuits and thus manipulating their dynamics by means of electrical currents is important for future developments.[4,5] Cylindrical symmetry gives raise to new possible magnetic configurations.[6–10] Typical nanowires investigated experimentally, with diameters above 50 nm, present two types of domain walls (DWs)[11–13]: vortex-antivortex (VAV) and the Bloch point (BP). The dynamics of both DWs is expected to be different from that of planar nanowires[14,15]. For example, DWs in cylindrical geometry have been predicted not to suffer from the Walker breakdown phenomenon, characteristic for the planar geometry, and thus potentially very high velocities, above 1000 m/s, have been theoretically predicted.[15,16] If these velocities can be achieved experimentally is still an open question.

Although the DW dynamics is well studied in planar magnetic nanowires (prepared by lithography), in cylindrical geometry only a scarce number of articles have reported experimental measurements. Ivanov et al[17] measured the motion of 3D domain walls by simultaneous application of field and current in bi-segmented Co/Ni nanowires, estimating DW velocity as few hundreds meter per second. Schöbitz et al[18] have observed current-induced domain wall motion in Ni-based nanowires by Magnetic Force Microscopy (MFM) and X-ray Magnetic Circular Dichroism (XMCD) combined with Photo-Emission Electron Microscopy (PEEM) measurements[18] estimating velocity up to 600 m/s. On the other hand, simulations show that during the current-induced dynamics the BP DW may be converted into the VAV domain wall, limiting its velocity. Additionally, the Oersted field was predicted to play an important role, being the source of DW transformation and dynamics, even without the direct action of the spin-transfer torques.[15,16]

Furthermore, if the BP DW velocities are found to be as high as theoretically predicted, the control of DW pinning will be an important aspect towards implementation of spintronics based on cylindrical nanowires. This may be achieved by creating special notches designed to stop their propagation. While the use of notches to pin DWs is well established in planar geometry[19,20], an efficient control of DW pinning under applied field in cylindrical nanowires has not yet been achieved.[21–23]

In this article we investigate the motion of DWs in Ni cylindrical nanowires with specially designed notches. To compare their effect with straight nanowires, we produced them only in one part of NW, leaving the other one free of defects. Significantly, while our experiment is successful in terms of nucleation, motion, and pinning, we also unexpectedly observe DW motion in the direction parallel to the current, i.e. against the electron flow. Our simulations including the spin-torque effects and the Oersted field assist in understanding the current-induced dynamics of DWs in the presence of notches. They show that DWs can be scattered from the notches and propagate in the opposite direction. Importantly, we identify a new DW pinning mechanism when the Oersted field with rotational sense opposite to initial BP DW can be a source of its pining outside the defect region.

**Experimental.** Cylindrical Ni nanowires with geometrical notches and high aspect ratio were grown by electrodeposition into the pores of anodic alumina membranes (Figs. S1(b)-Supplementary Information).[24] They were removed from alumina membranes by chemical etching and deposited onto a Si substrate. Finally, they were contacted with Au electrodes to allow the injection of electric current (Figs. S1(c-d)-Supplementary Information). More details about the fabrication and contacting procedures are given in the Supplementary Information. The SEM images of a contacted nanowire with the main diameter of about 100 nm and 13 μm length is displayed in Fig. 1. The geometry is schematically shown in the top panel. The close-up SEM images of Fig. 1 (b)-(c) correspond to the marked green and orange areas in Fig. 1 (a). The nanowire shows modulations/notches at the left side (area marked in green) and a uniform cylindrical geometry at the right side (area marked in orange).

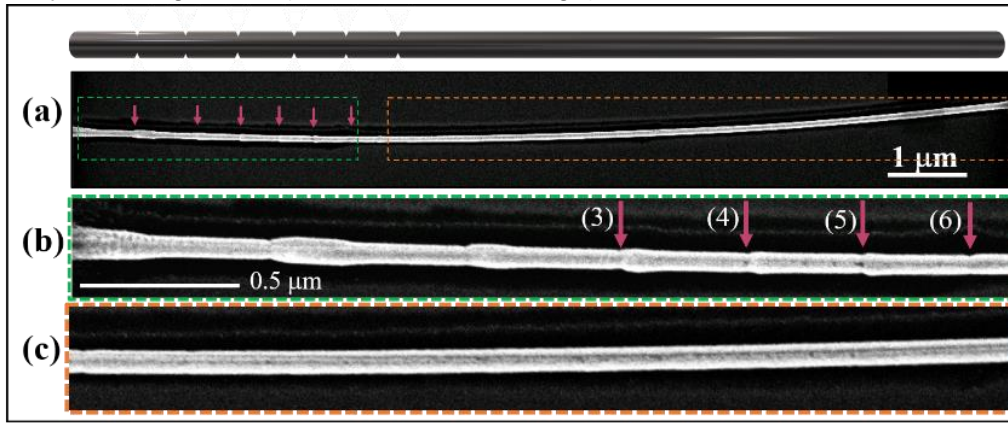

**Figure 1**. Schematic view of a contacted Ni nanowire (top panel). (a) SEM image of a contacted Ni nanowire with notches, marked by arrows, at the left-side end, (b) close-up SEM image of the green marked area in (a), (c) close-up SEM image of the orange marked area in (a).

Figure 2 presents XMCD-PEEM images of the same nanowire (NW) acquired at the Ni L3-edge with the X-ray incidence at about 45° to the nanowire axis, i.e., sensitive to both parallel and perpendicular magnetization components with respect to the NW axis. Fig. 2 (a) shows the NW in a single domain magnetic state as indicated by the uniform dark/bright XMCD contrast in the NW and its shadow[25], respectively. In the following, we will refer to the contrast of the NW itself when we write "dark" or "bright" contrast.

In Fig. 2 (b) the NW is imaged after applying a current pulse of $1.1 \times 10^{12}$ A/m$^2$ for 8 ns, with left polarity, that is large enough to break the magnetization into a multidomain state, as seen from the XMCD contrast, dark/bright/dark/bright/dark. There are five magnetic domains oriented antiparallel/parallel/antiparallel/parallel/antiparallel to the polarization vector (i.e. along the NW axis with three DWs pinned at the notches labelled (3), (4) and (6) and one DW pinned in the region of the NW with uniform diameter.

After the application of a second current pulse to the magnetic state shown in Fig. 2(b) with the same polarity but smaller amplitude, $1.5 \times 10^{11}$ A/m$^2$, the state in Fig. 2(c) is obtained. The magnetic images suggest that the DW initially pinned at the notch (6) has displaced towards the right, i.e., parallel to the electron flow (note that no pinning center is present at that point). Concerning the DWs initially pinned at notches (3) and (4), while the natural direction of motion is that the one pinned at the position (3) would displace to the right, the imaging does not allow us to exclude the situation in which the DW (4) would propagate to the left, i.e. opposite to the electron flow. The overall result is that only one main domain wall is visible in Fig.2(c). A further current pulse of $2.3 \times 10^{11}$ A/m$^2$ with the same polarity, applied to the magnetic state imaged in Fig. 2(c), pushes the DW from left to the right, finally developing into a single domain state (Fig. 2(d)).

Although we have only one data set for the current density of $2.3 \times 10^{11}$ A/m$^2$ where the DW has propagated outside the NW, we can estimate the lower bound of the DW velocity. Note that we have not seen additional DW nucleation at this current density. Neither we expect a depinning of an additional DW from the right end of the NW since the pinning there is strong and even in an ideal case cannot take place without an applied field.[26] Additionally the ends of the NW are kept at 300K and thus thermal depinning of DWs can be excluded. An estimation of the DW velocity from Fig. 2 (d) for a current pulse of $2.3 \times 10^{11}$ A/m$^2$, applied for 8 ns, gives rise to a value above 1000 m/s, in agreement with theoretical predictions[15,16,27] and higher than previously reported experimental values.[17,18]

Domain walls were nucleated again by applying a high-amplitude current pulse, $1.35 \times 10^{12}$ A/m$^2$ in the opposite sense. The resulting state, visible in Fig. 2 (e) has three domains and two DWs, one pinned at notch (3) and another at the left of notch (6). Then a sequence of low current pulses ($1.5 \times 10^{11}$ A/m$^2$) also with opposite sense to the initial ones was applied to study DW propagation. In this case, both DWs seem to displace along the NW, parallel to the applied current. In Fig. 2(f) the left-side DW is pinned at notch (6) and the right-side one has stopped at the right of notch (6) - outside but close to it. Upon application of a further identical current pulse, the NW saturates, as can be seen from the uniform bright contrast in Fig. 2 (g).

Additional information for the behavior of the DW movement is presented in Figs. S2 (b)-(c). In Fig. S2 (b) the NW is imaged before injecting a current pulse. Here we observed three magnetic domains oriented parallel/antiparallel/parallel to the polarization vector (i.e. along the NW axis). By injecting an intermediate current pulse of $5 \times 10^{11}$ A/m$^2$, the DWs, one pinned at the notch (6) and the other, outside of the notch (right-side end of the NW) move along the current, parallel to it, and the NW gets saturated, as can be observed from the uniform dark contrast in Fig. S2(c). Note that at these intermediate current densities, thermal depinning can be expected (see below).

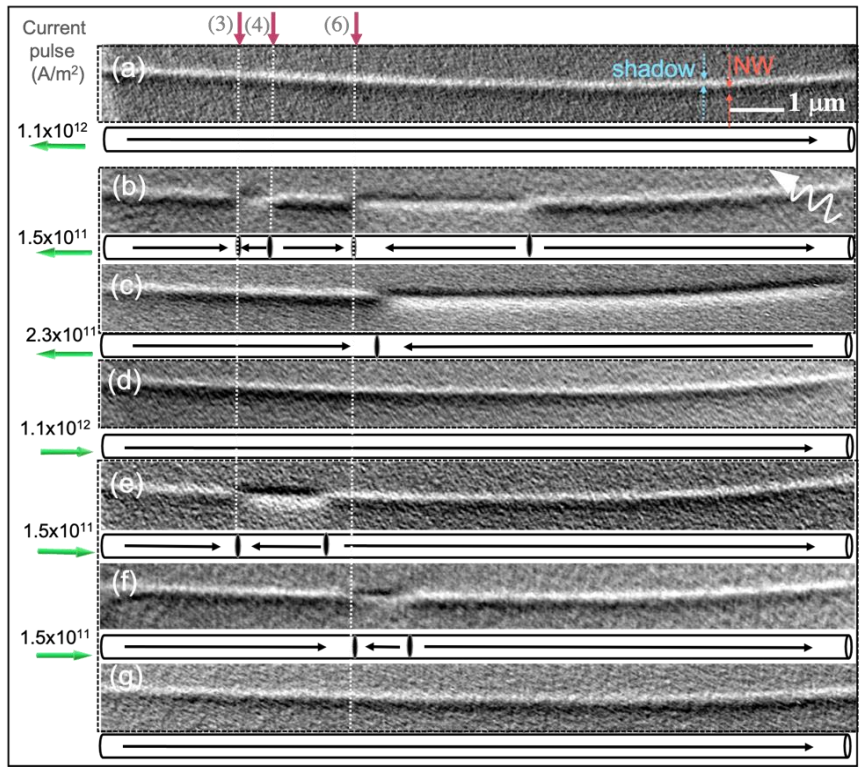

**Figure 2**. A sequence of XMCD-PEEM images taken after current pulses with different intensities and polarities (indicated by the green arrows) were applied along the contacted Ni modulated nanowire. a) XMCD-PEEM image of Ni NW presenting a single domain state. (b)-(g) XMCD-PEEM images of Ni NW taken after current pulses of 8 ns with different amplitudes and polarity, as represented in green color in the figure, were applied along its length. The red arrows and the white dashed line mark the different pinning centers produced by the notches. The graphical illustrations below each PEEM image represent the magnetic configurations of the NW.

To summarize the effect of the current, we observe that (i) Large current densities produce DW nucleation which appeared pinned after the pulse is switched off (Fig. 2 (b), (e)). (ii) Smaller currents move DWs. The results show DW dynamics with a lower current density than predicted in cylindrical nanowires with smaller diameter for transverse domain walls.[15] We attribute this fact to higher Bloch point domain wall mobility as compared to the transverse one.[16,18] While the largest DW propagation occurs against the current direction (i.e. along the electron flow as is expected and previously reported[17]), some propagation along the current is also observed. (iii) Some domain walls are stopped at notches (particularly notch (6)). However, others are stopped at different inter-notch places and their positions are not reproducible. This suggests a new pinning mechanism which cannot be fully ascribed to microstructural defects and modulations. Note that the DW propagation in the direction of the applied current as well as the pinning after the application of low-amplitude current pulses takes place in the left part of the NW, i.e. where the notches are present, while in the right part, without notches, the propagation is smooth and always against the current. (iv) For intermediate current pulses, the domain walls are moving both along and against the current direction. This happens also in the right part of the NW presented in Fig. S2 (c) (for much higher current densities) which we attribute to thermal domain wall depinning. All of the above indicates an important role of the notches in current-induced domain wall dynamics.

**Simulations:**

To understand the experimental behavior, we first assess the temperature rise due to the Joule heating by the current pulse of realistic temporal shape (more details in

Methods and Supplementary Information) and a maximum current density at 8 ns. The simulations show that temperature reaches maximum value at the timescale of 9 ns and decays to the ambient in the ca. 40-50 ns timescale. Figure 3 (a) presents the calculated maximum temperature distribution during the application of current along the direction of the axis in a single Ni NW showing that its maximum is reached at about 300 nm from the NW ends. Note that in simulations the temperature is kept at 300 K there due to the presence of contacts. Our modelling shows that the current pulses with magnitudes higher than $8\times10^{11}$ A/m$^2$ raise the temperature in the middle of the NW above the Curie temperature (for Ni, $T_c$=628K[28]). For a current pulse of $3\times10^{11}$ A/m$^2$ the temperature increase is around 30 K while for $1\times10^{11}$ A/m$^2$ it is only a few K.

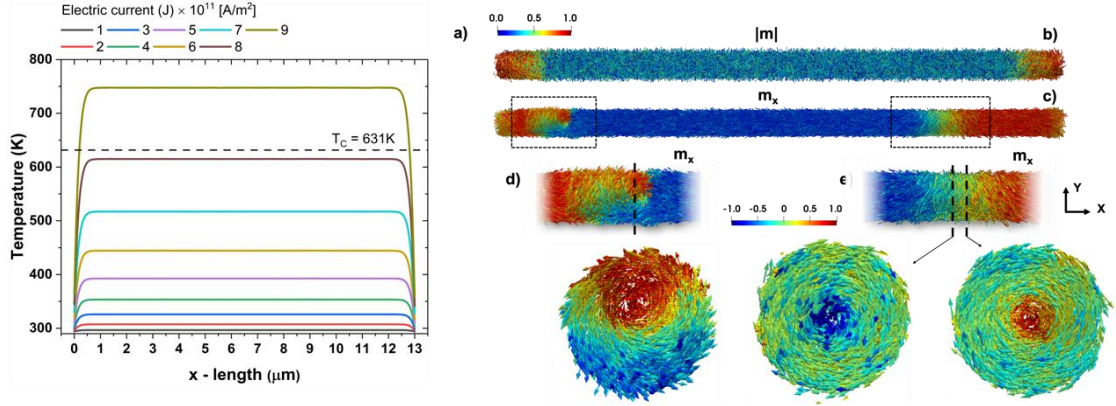

**Figure 3**. (a) Simulated temperature profile along the nanowire for different values of the current densities. (b-c) Results of thermal micromagnetic modelling showing the magnetization distribution in the wire colored by the reduced magnetization |m| (b) and the axial component $m_x$ (c) during (b) and after the pulse (c) for a temperature profile of 650 K maximum temperature. The images (d) and (e) show nucleated DWs of VAV (d) and BP type (e). The cross-section images of each domain wall are presented in the bottom panel.

To model the thermal nucleation with high current densities, we use a micromagnetic framework based on the Landau-Lifshitz-Bloch equation[29] with thermal fluctuations[30]. Importantly, this micromagnetics is a valid approach close and above the Curie temperature, $T_c$. The assumed temperature profile follows the one obtained by the COMSOL Multiphysics® software at 9 ns and lasts 50 ns. Our simulations show that the current pulses with the magnitudes corresponding to the heating below $T_c$ do not lead to any DW nucleation, neither during the heat pulse nor after it, speaking in favor of a large energy barrier for this process. Figure 3(b) shows the modulus of the reduced magnetization during the pulse when the maximum temperature of 650 K (higher than Tc) was reached. The wire presents a nearly null magnetization in its center while still being ferromagnetic at the edges due to the cooling effects of the electrodes.

Figure 3(c) shows the result of nucleation after a temperature profile with the maximum temperature of 650 K was applied. Importantly, the nucleation occurs after the temperature pulse, i.e. during cooling down of the magnetic system. The results show a pair of nucleated DWs: one of the VAV type (left in Fig.3 (c), augmented image in Fig.3 (d)), and another - of the BP type (right in Fig.3 (c) and augmented image in Fig.3 (e)). In the simulated system, these DWs propagate in the opposite directions in the absence of any external stimuli. In a realistic situation, they may be pinned at some pinning centers such as magnetic notches and defects. Note that the type of nucleated DWs is arbitrary and will vary in simulations with different temperature disorder.

We turn now to the computational results of magnetization dynamics at low current densities, considering the temperature fluctuations less relevant (more details in

Supplementary Information). The geometry represents a cylindrical NW of 100 nm in diameter and several microns in length with three notches and a pre-nucleated BP DW (initially pinned at the central notch for convenience). Since previous reports indicate that the VAV DW is transformed to the BP DW during the dynamics[31], we consider only the last case. The current interacts with magnetization through the Zhang-Li spin transfer torque and the associated Oersted field.[32] Note that two types of the BP domain walls can exist: one with the rotational sense parallel to the Oersted field direction (called here "good" chirality) and the other one - with the rotational sense antiparallel to it ("bad" chirality BP DW). The BP DW is followed by the intersection of three iso-surfaces defined by $m_x=0$, $m_y=0$, $m_z=0$.

Our modelling results of DW dynamics with the current switched on are presented in Figs. 4 and 5. It is worth to mention that the XMCD-PEEM images are done in the stationary situation after the current is switched off and DW is stopped while the simulations allow to understand dynamical processes during the current application. The simulations show that in all cases the DW dynamics is finished at 8 ns (either because a fast DW propagated outside the NW or it stopped the propagation due to pinning), thus the pictures at 8 ns will reflect the stationary situation.

The modelling results indicate that under considered current densities ($1\times10^{11}$ - $4\times10^{11}$ A/m$^2$) the DW always gets unpinned from the central notch. The large-distance propagation typically occurs in the NW axis following the direction opposite to the direction of the current (indicated by dark blue arrows on the top of the graphs), i.e. in the direction of the electron flow. This result is general for negative (left) and positive (right) applied currents and explains the main direction of the propagation in the experimental results. It also corresponds to the sign of the non-adiabatic spin-transfer torque reflecting the direction of the electron flow. Note the complex dynamical DW structure (indicated by the iso-surface with $m_x=0$, reddish region) which is dynamically changing during the propagation (called "butterfly" DW). An important remark is that only a BP DW with "good" chirality driven by intermediate current densities can cross the whole NW without pinning, see Fig. 4 (a). For all other situations we observe a pinning process either near the left notch (small current densities and "good" BP chirality, Fig.4 (b)) or near the right notch (small current densities and "bad" BP chirality, Fig.5 (b)) or even in the middle constriction (large current densities and "bad" BP chirality, Fig.5 (a)).

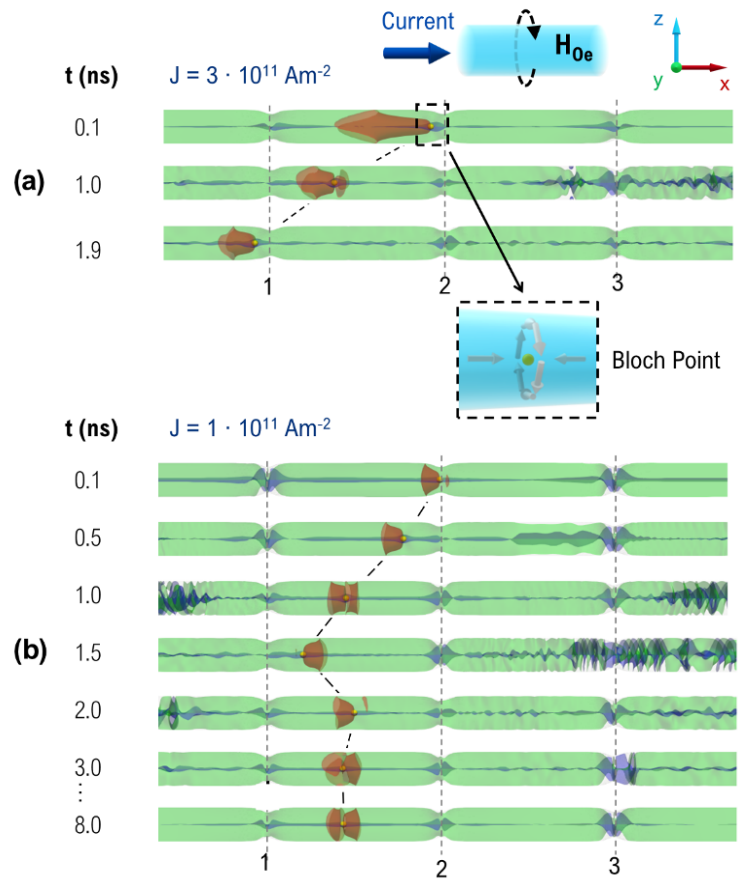

**Figure 4**. Dynamics of a Bloch Point domain wall with 'good' chirality under two applied current densities in a nanowire with three geometrical notches labelled 1-3. The dark blue arrow above the images indicate the current direction while the dashed black arrow – the direction of the Oersted field ($H_{Oe}$). The BP DW characteristics (close-up image) are presented in the inset (dashed square), i.e. initially DW has the head-to-head structure with counter-clockwise rotational sense coincided with that of the Oersted field. The iso-surfaces of the x-, y-and z- zero components of magnetization are coloured in red, green, blue and the yellow sphere indicates the position of the Bloch Point. (a) Bloch Point with 'good' chirality propagates in the direction opposite to the driving spin-polarized current for intermediate current density $J=3\times10^{11}$ A/m. (b) A low current density $J=1\times10^{11}$ A/m produces domain wall recoil from the pinning center and temporal propagation in the opposite direction.

An inspection of the pinning process for low positive currents ("good" DW chirality) in Figs 4 (b) indicates that the BP is first unpinned from the middle notch but then recoils (bounces) when approaching the neighbouring notch, which leads to its propagation in the opposite direction, and finally it stops at an intermediate position in the segment in few nanoseconds. We attribute this recoil to the presence of small chiral structures (not spatially resolvable by PEEM) created at the notches. Due to magnetostatic energy minimization, together with the action of a different local Oersted field, the magnetization at the notch acquires certain circulations (visible as blueish regions in Figs.4-5). These structures interact with the DW and are responsible for its repulsion from the notch. Note also a strong emission of spin-waves from NW ends and notches, which also interact with the DW.

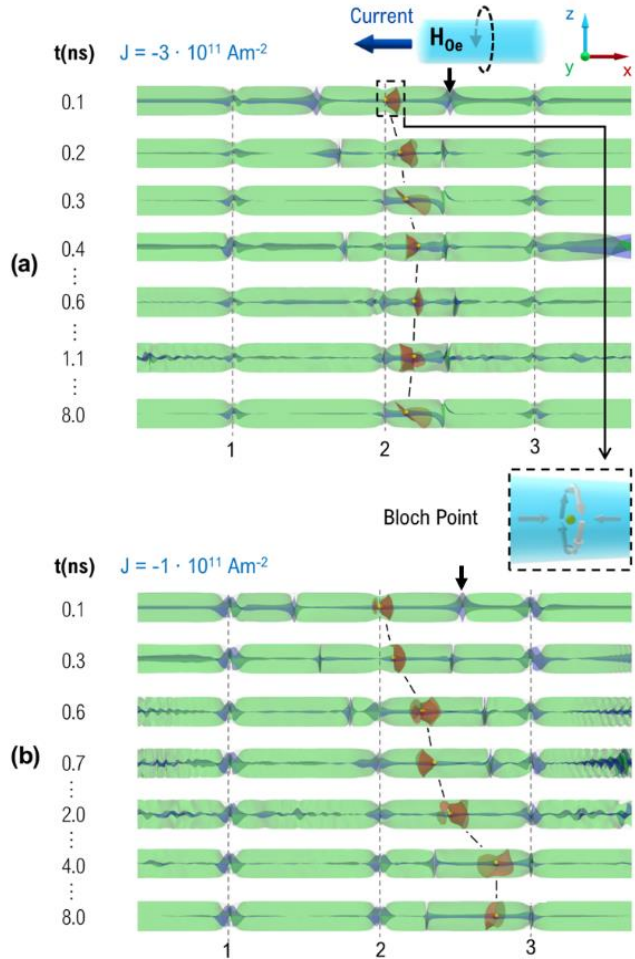

**Figure 5**. Dynamics of Bloch Point (marked by the red arrow) with 'bad' chirality under two applied current densities in a nanowire with three geometrical constrictions. The dark blue arrow above the images indicate the current direction while the dashed black arrow – the direction of the Oersted field. The BP DW characteristics (close-up image) are presented in the inset (dashed square), i.e. initially DW has the head-to-head structure with counter-clockwise rotational sense opposite to that of the Oersted field Red, green, blue colours represent isosurface of $m_x=0$, $m_y=0$, $m_z=0$, respectively (a) Bloch Point with 'bad' chirality is pinned by the effect of the oersted field produced by spin-polarized current for intermediate current density $J=-3\times10^{11}$ A/m. The black arrows indicate quasi-domain walls created by the encounter of two chirality: from the BP DW and that created by the Oersted field (b) At low current density $J=-1\times10^{11}$ A/m domain wall propagates in the direction opposite to the driving spin-polarized current until it is bounced from the notch.

The same DW is of the "bad" chirality type when the current is applied in the opposite direction. These DWs may experience a chirality switching,[31,33] however, we did not observe it in these particular cases (see results in Fig.5). For negative currents (oriented to the left), DW first propagates to the right and is pinned close to the right notch. What is striking is that for higher currents this DW gets stopped near the initial pinning site but at a position farther from the notch. This effect is due to the Oersted field creating an azimuthal magnetization component with the direction opposite to that of the BP DW, and thus there appears one more quasi-domain wall, (indicated by black arrows) where the two chirality meet each other. This quasi-domain wall, which is induced by the

Oersted field, is quite robust and topologically protected (since in order to annihilate it one would need to unwind one of the chiralities) and is a source of pinning for the initial Bloch Point wall. A similar quasi-domain wall is created at the left side from the original domain wall. Both together produce its pinning.

In summary, the simulations show that BP DW with "good" chirality move under current, but can be reflected from notches, moving more efficiently at higher current density. In contrast, results are different for the "bad" chirality BP DW: while at low current densities an equivalent motion as for the "good" chirality is observed, at higher current densities a pinning by the Oersted field sets in, cancelling any motion under these conditions. Thus, DW mobility is limited by the "bad" chirality only at high current density, but is not affected at low current density.

In conclusion, by combined theoretical and experimental studies we have demonstrated the possibility to move Bloch point domain walls by means of current pulses in modulated Ni nanowires. First domain walls were nucleated by current pulses with high amplitude which raise the temperature in the middle part of the nanowire above the Curie temperature and provide thermally induced stochastic nucleation. When the system cools down, these domain wall appeared pinned at the notches. For smaller currents, domain walls can be moved. As expected, the largest domain wall propagation occurs in the direction antiparallel to the applied current direction (parallel to the electron flow) and the pinning occurs at the notches. Unexpectedly, we also observe short-distance propagation in the direction parallel to the current as well as pinning of domain walls outside the notches. Our dynamical simulations help to ascertain the origin of these effects. We have identified two different pinning mechanism for domain walls in cylindrical nanowires, one related to the effect of notches themselves and another induced by the Oersted field which creates quasi-domain walls of a chiral nature. As for the first mechanism, it seems that the pinning is additionally assisted by Oersted field and a more efficient control can be expected than that with external field. For the latter case, to our knowledge, this is the first time that this mechanism is reported. Importantly, it is not related to the presence of notches and will limit the "bad chirality" domain wall propagation in spintronics devices. Finally, we cannot exclude the influence of thermal depinning mechanisms for intermediate current densities.

An important finding is our estimation of the domain wall velocity in the region free of pinning centres, as high as theoretically predicted, i.e., of the order of 1 km/s, and higher than previously observed experimentally. We also report lower current densities which move DWs in relation to previously reported results. We believe our results to be of high significance for further developments of spintronic applications based on cylindrical magnetic nanowires.

**Methods.**
*Nanowires fabrication.* Ni nanowires with imprinted notches along their length were produced by electrodeposition in anodic aluminum oxide (AAO) templates.[24,34–36] The modulated pores of AAO templates are produced by pulsed hard anodization (Fig. S1 (a)) in oxalic aqueous solution. Ni nanowires were electrodeposition the pores of fabricated AAO templates by electrodeposition in a three-electrode cell using a single Watts-type bath containing 0.38 M $NiSO_4 \cdot 7H_2O$, 0.08 M $NiCl_2 \cdot 6H_2O$ and 0.16 M $H_3BO_3$ at a constant voltage of 1V for about 40 minutes.
The total length of the nanowires in the membrane is about 28 □m. The center-to-center inter-pore distance is kept constant at 320 nm (see Fig. S1 (b)).

After the alumina template was dissolved by chemical etching the nanowire were dispersed on a Si/SiO$_2$ substrate and electrically contacted. More details about the sample preparation are presented in Supplementary information file.

*XMCD-PEEM measurements.* The XMCD–PEEM measurements were performed at the CIRCE beamline of the ALBA Synchrotron Light Facility (Barcelona, Spain) using an ELMITEC LEEM III instrument with energy analyzer.[37] The samples are illuminated with circularly polarized X-rays at a grazing angle of 16° with respect to the surface. The emitted photoelectrons (low energy secondary electron with ca. 1 eV kinetic energy) used to form the surface image are proportional to the X-ray absorption coefficient and thus the element-specific magnetic domain configuration is given by the pixel-wise asymmetry of two PEEM images sequentially recorded with left-and right-handed circularly polarized light.[37,38] Sample holders with integrated printed circuit board (PCB)[39] were used for the injection of current pulses into the NW, with ultrasonic wire bonds as connections to the gold electrode of the sample.

*Thermal simulations* of individual modulated Ni nanowires have been carried out using COMSOL Multiphysics® software to determine the temperature in the system under the influence of the current pulses with a rise time of 8 ns and total duration of 15 ns. We assume a 13 μm Ni nanowire slightly buried in SiO$_2$ substrate and Cr/Au contacts maintained at 300K. More details in Supplementary Information.

*Micromagnetic simulations.* Thermal nucleation close and above the Curie temperature with spatial temperature profiles calculated from COMSOL Multiphysics® simulations have been investigated using home-made micromagnetic code based on the Landau-Lifshitz-Bloch approach. The current pulse duration up to 50 ns were considered to emulate the heat accumulation in the system. The non-thermal magnetization dynamics simulations are done using the Landau-Lifshitz-Gilbert micromagnetics with Zhang-Li spin-transfer torque as implemented in the mumax program.[40] DW dynamics was investigated under a spin-polarized current with polarization P=0.5, non-adiabaticity ξ=0.1 and damping 0.02 in a Ni nanowire with standard parameters of Ni used in previous studies.[41] (more details in Supplementary Information). Importantly, here the Oersted field takes into account the reduced current density at the modulations.

ASSOCIATED CONTENT

**Supplementary Information.** The Supporting Information is available free of charge on the ACS Publications website

Details of nanowires fabrication and preparation of electrical contacts, micromagnetic simulations of Joule heating and domain wall dynamics.

AUTHOR INFORMATION

Corresponding Author: Cristina Bran, email: cristina.bran@icmm.csic.es

**Acknowledgements:**

This work was supported by the grants PID2019-108075RB-C31 funded by Ministry of Science and Innovation MCIN/AEI/ 10.13039/501100011033 and S2018/NMT-4321 NANOMAGCOST-CM funded by the Government of Madrid Region, Spain. We acknowledge the service from the MiNa Laboratory at IMN, and funding from CM (project SpaceTec, S2013/ICE2822), MINECO (project CSIC13-4E-1794) and EU (FEDER, FSE).

# Supplementary Information

# Domain wall propagation and pinning induced by current pulses in cylindrical modulated nanowires

C.Bran[1], J.A. Fernandez-Roldan[2], J. A. Moreno[3], A. Fraile Rodríguez[4,5], R. P. del Real[1], A. Asenjo[1], E. Saugar[1], J. Marqués-Marchán[1], H. Mohammed[3], M. Foerster[6], L. Aballe[6], J. Kosel[3,7], M. Vazquez[1], O. Chubykalo-Fesenko[1]

[1] Instituto de Ciencia de Materiales de Madrid, 28049 Madrid, Spain

[2] Helmholtz-Zentrum Dresden-Rossendorf e.V., Institute of Ion Beam Physics and Materials Research, Bautzner Landstrasse 400, 01328 Dresden, Germany

[3] King Abdullah University of Science and Technology, Computer Electrical and Mathematical Science and Engineering, Thuwal 23955-6900, Saudi Arabia.

[5] Departament de Física de la Matèria Condensada, Universitat de Barcelona, Barcelona, 08028, Spain

Institut de Nanociencia i Nanotecnologia (IN2UB), Universitat de Barcelona, Barcelona, 08028, Spain

[6] ALBA Synchrotron Light Facility, CELLS, Barcelona, 08290, Spain

[7] Sensor Systems Division, Silicon Austria Labs, Villach 9524, Austria

**Sample preparation.** Modulated pores are produced by pulsed hard anodization in oxalic aqueous solution. By tuning the electrochemical parameters periodically geometrical modulations were imprinted along the nanopores length. The templates with modulated pores were obtained in oxalic aqueous solution (0.3M) containing 5 vol.% ethanol at a constant temperature of 0 ºC. During anodization, a constant voltage of 80 V was first applied for 400 s to produce a protective aluminum oxide layer at the surface of the disc which avoids breaking or burning effects during subsequent pulsed hard anodization (Fig.S1a-(I)). In the second step, the voltage was steadily increased (0.08 V/s) up to 130V and kept constant for 400s, which ensures the alignment of the nanochannels (Fig.1(a)-(II)). Nanopores with tailored periodical diameter modulations were produced in step III by applying pulses of 130V and 100V for 5 and 50 s, respectively (Fig. S1-(III)). The pulses were repeated 50 times to obtain a total length of the modulated pores of few tens of microns (Fig.S1 (b)). After an Au nanolayer was sputtered on the back side of the template, Ni magnetic nanowires (Fig. S1 (b)) were grown into the cylindrical modulated pores by electrodeposition taking the shape of already modulated AAO pores [1]. The total length of the nanowires in the membrane is about 28 µm. The cylindrical wires are formed by 500-800 nm long segments with diameter of about 120 nm which are periodically separated by small areas where the diameter is decreased at about 90 nm forming geometrical notches along the nanowire´s length (Fig S1 (b)-inset marked in blue). The center-to-center inter-pore distance is kept constant at 320 nm. The variation in the current during anodization (Fig. S1(a)-inset) produces the bottle necks in most of the segments (Fig. S1 (b)-inset and Fig. 1 (b)).

**Electrical contacts.** The substrates used to contact the single NWs (Fig. S1 (c)-(d)) are 1x1cm² Si/SiO$_x$ with 90 ± 5 nm oxide thickness. These were cleaned and then baked at 60 °C for five minutes in a hotplate in order to remove adsorbed molecules. To prepare a device, a 2µl drop of ethanol containing NWs was released over a substrate right after it was removed from the hotplate and allowed to dry over a permanent magnet. Using a scanning electron microscope (SEM) equipped with a focused ion beam (FIB) this substrate containing NWs was inspected in search for an isolated NW near the center of the substrate. Such NW was signaled by etching marks parallel to its long axis (100 μm away). In order to pattern the contacts over the NW, a two-step lithographic process was used in which LOR5B and AZ5214 photoresists were used sequentially to create an undercut and ease lift-off [2].

To deposit the metallic contacts over the developed samples, a sputtering setup with load-lock plasma etching capabilities was used. Under a base pressure of $10^{-7}$ Torr, the samples were etched in the load-lock in order to remove the naturally formed oxide layer of the NW and without removing the vacuum, they were transferred to the sputter chamber where 10 nm of chromium and 180 nm of gold were deposited as electrical contacts. Finally, the photoresists (and excess sputtered materials) were lifted-off by submerging in two sequential baths of preheated Remover PG at 60 °C for at least four hours, dipped in IPA, DI water and nitrogen blow dry.

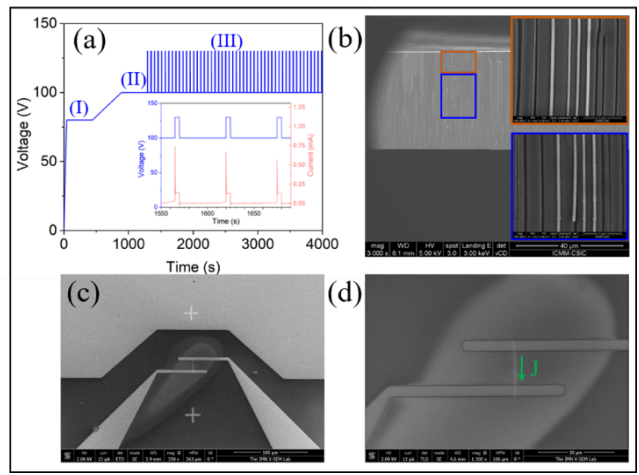

**Figure S1**. (a) Voltage–time transients of the anodization, including the first anodization step (I), ramping of voltage and keep it constant (II) and the applied pulses (III). The inset shows a close look of both, applied voltage and current transients, during the pulsed anodization (b) SEM image of Ni modulated nanowire embedded in alumina template, (c) Cr/Au electrical contacts, (d) contacted Ni modulated nanowire.

Figure S2 presents the x-ray absorption spectroscopy (XAS) image and XMCD-PEEM images of the Ni modulated nanowire. In (b) the NW is imaged in the as-prepared state showing two domain walls, one of them pinned at the modulation (6), the other one outside the modulation's region. After an intermediate current pulse of $5\times10^{11}$ A/m$^2$ was applied, they move along the current and the nanowire gets saturated as can be observed from the uniform dark contrast in Fig. S2(c).

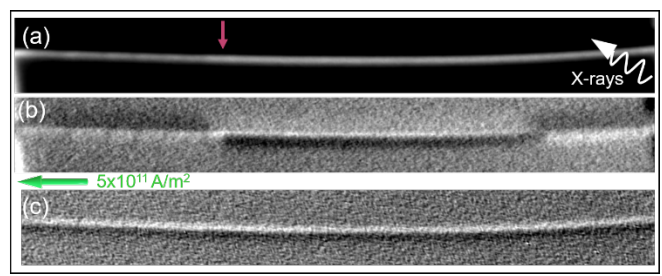

**Figure S2**. (a) XAS and XMCD-PEEM image of Ni nanowire in as-prepared state (b) and (c) after an intermediate current pulse of $5\times10^{11}$ A/m$^2$ was applied.

## Micromagnetic modelling

*a) Modelling Joule heating in an individual straight Ni nanowire:*

The temperature distribution along the NW axis for different electrical current density values was evaluated with COMSOL Multiphysics® [3] simulations. The geometry of the simulated device was the same as the experimental set-up. The electrical conductivity $\sigma_e$ as a function of the temperature $T$ of the nickel NW was extracted from [4], and the values above room temperature were extrapolated following the equation $\sigma_e = A + B/T + C/T^2 + D/T^3$, where $A = (-1.6 \pm 0.7) \cdot 10^5 \, \Omega^{-1} \cdot m^{-1}$, $B = (1.22 \pm 0.04) \cdot 10^9 \, K \cdot \Omega^{-1} \cdot m^{-1}$, $C = (-1 \pm 7) \cdot 10^9 \, K^2 \cdot \Omega^{-1} \cdot m^{-1}$, $D = (4.8 \pm 0.4) \cdot 10^{11} \, K^3 \cdot \Omega^{-1} \cdot m^{-1}$. The thermal conductivity $k$ of the NW was extracted from the electrical conductivity using the Wiedemann-Franz law $k/\sigma_e = L \cdot T$, where $L$ is the Lorenz number and is set to $2.8 \cdot 10^{-8} \, W\Omega/deg^2$ [4]. The rest of parameters were extracted from COMSOL library.

Electrical current pulses with amplitudes $j_0$ from 1 x 10$^{11}$ to 1 x 10$^{12}$ A/m² follow the equations $j_{rise} = j_0(1 - e^{-t/2})$ and $j_{fall} = j_0(e^{-(t-t_p+\tau)/2})$, where $j_{rise}$ and $j_{fall}$ are the rise and fall curves of the pulse, t is the time, $t_p$ = 15 ns is the total pulse duration, and $\tau$ = 7 ns the rise time. This shape of the pulse (Figure S3) is chosen in order to assume the deformation of the squared pulse [5].

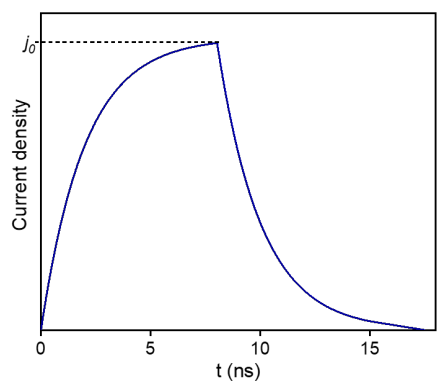

**Figure S3**. Electrical current pulse used in COMSOL simulations.

*b) Current-induced domain wall dynamics and pinning in modulated nanowires.*

In order to understand the magnetization processes observed in the XMCD-PEEM experiment, we have investigated the dynamics of a domain wall in a Ni nanowire with a total length 2L = 5 µm, a diameter of D$_o$=100 nm, and periodical notches given by a nanowire width

$$D(z) = D_o - \Delta \left( \text{sech}\left(\frac{x}{w}\right) + \text{sech}\left(\frac{x+L}{w}\right) + \text{sech}\left(\frac{x-L}{w}\right) \right) \quad (1)$$

as a function of the axial coordinate $x \in (-L, L)$. $\Delta = 12$ nm and w=16 nm are adequate values to approach the dimensions observed in the SEM images. In this approach the notches, labelled from 1 to 3, are spaced by 500 nm as depicted in Fig S4(a). The magnetization is colored by its axial component.

The implementation is carried out with the finite different package mumax3 [6] (version 3.10) and cell size 1x1 nm. We used typical parameters of Ni in Table 1. Magnetic poles have been removed from the ends of the nanowire to mimic long wire conditions in the demagnetizing field. We introduced the pre-calculated Oersted field for a current flow through a nanowire with profile $D(z)$ in Eq. 1. The spin-polarized current has a polarization P=0.5, non-adiabaticity ξ=0.1 and damping 0.02.

Table 1. Material parameters and crystal structural in micromagnetic modelling.

| Material | $\mu_o M_s$ (T) | $A_{ex}$ (pJ/m) | Crystal Symmetry | $K_1$ (kJm$^{-3}$) | Magnetization Easy Axis (e.a.) |
|---|---|---|---|---|---|
| Ni (111) [7] | 0.61 | 3.4 | Cubic | -4.8 | [111] crystal lattice direction parallel to the nanowire axis |

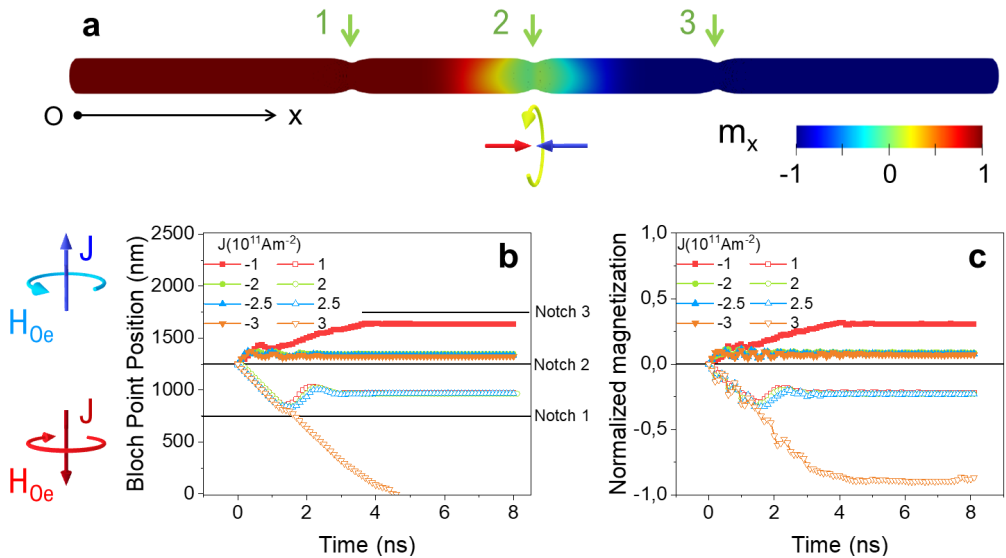

**Figure S4**. Initial Bloch Point Wall configuration in a Ni nanowire with three notches (a). The green arrows indicate the position of the notches from 1 to 3, the schematic view below the wire represents the Bloch Point wall structure. (b)The position of the Bloch point as a function of time during propagation under current in the nanowire for various current density values (c) Normalized axial magnetization component as a function of time for the same current density values. A vast panorama of dynamic and pinning processes which cannot be accurately interpreted from the 'wavy' magnetization dynamics, reflecting spin-wave emission.

We start with a pre-nucleated head-to-head Bloch Point Domain Wall (BP DW) at the notch 2 as indicated in Figure S4(a). In this example the direction of the Oersted field for positive currents has the same sense as the azimuthal component of the magnetization of the domain wall, i.e., the BP DW has a 'good chirality'.

The dynamics start immediately with the switching ON of a current pulse of density J and its Oersted field. It is important to bear in mind that the temporal dependence of the position of the BP DW and the mean magnetization in the wire in Figures S4(b-c) are strictly not proportional one to another mainly due to the emission of spin waves during the dynamics which is observed in Fig S4(c). In regard to the position of the BP DW for positive currents in Figure S4(b), we observe that the propagation of the Bloch Point occurs in the direction opposite to the direction of current (indicated by blue arrow of J on the left side of the graph). This result is general for negative and positive currents in Figures S4(b-c).

The BP DW can pass the notch only for high current densities (>3·10$^{12}$ A/m$^2$). For lower current density the BP DW remains pinned at some position. Strikingly, the pinning position is the same for every pinned Bloch Point irrespectively of the value of the current density.